\documentclass[twoside,a4paper,11pt]{article}
\usepackage{irrj2e}
\usepackage{lastpage}
\usepackage{hyperref}
\usepackage{url}
\usepackage[utf8]{inputenc}
\usepackage{booktabs}
\usepackage{amsmath}
\usepackage{amsfonts}
\usepackage{amssymb}
\usepackage[inline]{enumitem}
\usepackage{enumitem}
\usepackage{mathbbol}
\usepackage[T1]{fontenc}
\usepackage{mathtools}
\usepackage{multirow}

\newcommand{\ie}{\emph{i.e.}}
\newcommand{\eg}{\emph{e.g.}}

\newcommand{\paperTitle}{Emancipatory Information Retrieval}

\irrjheading{1}{2025}{313-\pageref{LastPage}}{9/25}{12/25}{10.54195/irrj.24531}
\ShortHeadings{\paperTitle}{Mitra}
 \firstpageno{313}
\title{\paperTitle}
\author{\name Bhaskar Mitra
  \email bhaskar.mitra@acm.org \\
  \addr Tiohtià:ke/Montréal, Canada}
\editor{Djoerd Hiemstra}

\begin{document}

\maketitle
\begin{abstract}
Our world today is facing a confluence of several mutually reinforcing crises each of which intersects with concerns of social justice and emancipation.
This paper is a provocation for the role of computer-mediated information access in our emancipatory struggles.
We define emancipatory information retrieval as the study and development of information access methods that challenge various forms of human oppression, and situates its activities within broader collective emancipatory praxis.
The term "emancipatory" here signifies the moral concerns of universal humanization of all peoples and the elimination of oppression to create the conditions under which we can collectively flourish.
To develop an emancipatory research agenda for information retrieval (IR), in this paper we speculate about the practices that the community can adopt, enumerate some of the projects that the field should undertake, and discuss provocations to spark new ideas and directions for research.
We challenge the field of IR research to embrace humanistic values and commit to universal emancipation and social justice.
We also invite scholars from fields such as human-computer interaction, information sciences, media studies, design, science and technology studies, social and political sciences, philosophy, law, environmental sciences, public health, educational sciences, as well as legal and policy experts, civil rights advocates, social justice activists and movement organizers, and artists to join us in realizing this transformation.
In this process, we must both imagine post-oppressive worlds, and reimagine the role of IR in that world and in the journey that leads us there.
\end{abstract}


\section{Introduction}
\label{sec:intro}

\begin{quote}
  ``The old world is dying, and the new world struggles to be born: now is the time of monsters.''
  \begin{flushright}
  -- Antonio Gramsci
  \end{flushright}
\end{quote}

\paragraph{Our world in crises.}
Our world today is facing a confluence of several mutually reinforcing crises pushing us universally, but notably not uniformly, towards precarity.
The recent COVID-19 pandemic did not only result in over seven million deaths~\citep{mathieu2024coronavirus} and disrupt everyday lives globally; it also laid bare and further reinscribed existing political, economic, and social hierarchies that are saliently colonial and racialized~\citep{papamichail2023reinscribing}, gendered~\citep{carli2020women, madgavkar2020covid}, casteist~\citep{mondal2024caste}, and ableist~\citep{lund2022ever}.
Likewise, the dangers of recent and impending climate catastrophes as well as the discourse around mitigative actions are inextricably intertwined with concerns of racial~\citep{sultana2024confronting, williams2022climate, patnaik2020racial, stefani2021indigenous}, gender~\citep{gender2016gender, women2022explainer}, and disability~\citep{cram2022cripping} justice.
Increasing global wealth and income inequality~\citep{chancel2022world} both mirrors historical dimensions of marginalization and is contributing to erosion of democracies worldwide~\citep{rau2025income}.
Globally, we are still reckoning with the ongoing legacies of settler-colonialism~\citep{wolfe2006settler} and present-day neocolonialism~\citep{nkrumah1965neo}, recontesting women's rights to bodily autonomy~\citep{beck2024reproductive}, and witnessing growing assaults on transgender people's right to exist~\citep{human2023epidemic, horbury2020empire}.

The range of crises we face in the twenty-first century requires us to reckon with multiple intersecting and co-constituting processes of hierarchical organization of our society and as a consequence with the systemic dehumanization, marginalization, and oppression of those designated to its bottom rung.
Therefore, any attempt at addressing the crises of our times must begin with recommitment to the principles of social justice and universal humanization of all peoples and support for emancipatory struggles against social, economic, political, cultural, religious, sexual, and bodily oppression.
This paper is a provocation for the role of computer-mediated information access in these struggles.
Our aim is to challenge the field of information retrieval (IR) research to reimagine itself as what it can (or, ought to) be if grounded in humanistic values and committed to universal emancipation and social justice, and to encourage the community to explicitly articulate and make conscious choices about the values and incentives that shapes our field of computing.
This is also an invitation to scholars from fields such as human-computer interaction (HCI), information sciences, media studies, design, science and technology studies (STS), social and political sciences, philosophy, law, environmental sciences, public health, educational sciences, as well as legal and policy experts, civil rights advocates, social justice activists and movement organizers, and artists to join us to collectively reimagine radically new sociotechnical futures and realize desired transformations.

\paragraph{The information reaches out.}
IR, the field of computing that is concerned with the design and study of information access systems, such as web search and recommender systems, has its roots in library science and linguistics~\citep{harman2019information}.
Early IR research was primarily concerned with the design of effective indexing mechanisms that enable fast pattern-based lookup of text from large corpora.
The modern-day successors of these systems, however, go far beyond serving the function of simple lookup.
Modern search systems observe and model collective behaviors of how their users interact with retrieved information to infer intent behind user queries, and to predict users' future needs and actions.
Some of these systems may proactively surface new information and breaking news, recommend commercial products and services, and even suggest activities and actions for the user to undertake.
Conversational systems, such ChatGPT,\footnote{\url{https://chat.openai.com/}} Microsoft Copilot,\footnote{\url{https://copilot.microsoft.com/}} and Google Gemini\footnote{\url{https://gemini.google.com/}}, may be entrusted to summarize retrieved information in response to users' search queries, effectively shifting the burden of assessing the relevance and trustworthiness of retrieved information and their contextual interpretation from the user to the statistical models underlying these systems.
These systems affect what information and perspectives receive exposure which shapes consumer behavior, political discourse, and culture~\citep{grimmelmann2008google, gillespie2019algorithmically, hallinan2016recommended}, and directly influence material outcomes for people~\citep{singh2018fairness, datta2014automated}.

Today's search and recommender systems are, therefore, much more than passive lookup tools.
These systems have become embedded in our online social discourse, both shaping and being shaped by our values, cultural identities, sociopolitical beliefs, and shared understanding of our environment and ourselves.
We could choose to see this transformation of information access systems from passive lookup tools to agents of influence as the information reaching out and permeating every aspect of our individual and social lives---akin to how \citet{grudin1990computer} described the transition of computers from applications in narrow specialist domains to their present-day ubiquitous presence in our lives as ``the computer reaches out''.

\medskip\noindent
Importantly, these systems and platforms may privilege certain perspectives and ways of knowing over others as forms of epistemic injustice~\citep{fricker2007epistemic} that entrench social hierarchies and intensify oppression.
Or, alternatively they could serve to raise up marginalized voices and challenge the status quo.
Access to information is critical to collective sense-making of our place and relationships in this world.
Therefore, it is unsurprising that throughout history authoritarian forces have tried to control what information is disseminated and how.
In both historical and ongoing conflicts, we can observe these struggles for control over both traditional media (\eg, newspapers, radio, and television) and online media (\ie, social media and other online information access platforms).
For the oppressed, they serve as medium for building shared understanding and solidarity necessary for social movements and collective action.
For the oppressor, they represent sites to apply force to censor information and perspectives and manipulate public opinion.
Information access platforms, therefore, represent sites of conflict between exploitation and liberation.
\emph{How should we then reflect on and reconcile the implications of IR research and development on our communities and our collective liberation?}
In this paper, we assert that the field of IR would benefit from critically reflecting on how our work relates to current sociopolitical orders and our possible sociotechnical futures, and argues for reimagining IR research grounded explicitly in the adoption of humanistic and emancipatory values and aspirations.


\section{What is emancipatory IR?}
\label{sec:what}
We define \emph{emancipatory IR} as the study and development of information access methods that challenge all forms of human oppression, and situates its activities within broader collective emancipatory praxis.
The term ``\emph{emancipatory}'' here signifies the moral concerns---central to this field of study---of universal humanization of all peoples and the elimination of oppression to create the conditions under which we can collectively flourish.
Historically, the term ``emancipation'' has taken on various meanings in context of diverging intellectual traditions~\citep{susen2015emancipation}.
It described the abolitionist struggles against slavery, but later came to signify a broader vision of dismantling all forms of structural oppression.
Examples of structural oppression in this context include colonialism, racism, patriarchy, casteism, transphobia, religious persecution, and ableism.
Emancipatory thinking in humanities have manifested in many forms, including Marxism, Critical Theory, feminism, and  decolonialism. 
Emancipatory IR aims to be informed by and incorporate these different theories and epistemic frameworks in the conception and design of information access methods and systems.

This framing recognizes information and access to information both as profoundly political, and enacts research and development activities---such as, formalization, theorization, design, experimentation, publishing, open sourcing, deployment, platform governance, and community building---in service of universal struggles against all forms of structural oppression.
It rejects the techno-deterministic premise  that there is a single pre-determined path forward for information access technology development, and challenges the community to employ reflexivity to uncover the deeply embedded values, incentives, and sociotechnical imaginaries that shape the field of IR.
It recognizes the agency of the field and its contributing members in shaping the world, while respecting that the challenges and necessary interventions are both fundamentally sociotechnical in nature.
Therefore, the theories and practices of this field must be co-developed with cross-disciplinary collaboration including with scholars from HCI, design, information sciences, media studies, STS, social and political sciences, philosophy, law, environmental sciences, public health, educational sciences, as well as with legal and policy experts, civil rights advocates, social justice activists and movement organizers, and artists, among others.

Emancipatory IR challenges us to employ technological research and development in service of dismantling hierarchies of oppression and challenge power.
It discourages non-performative academic gaze, and urges this research to be situated in movement building, and calls for recognizing the role of movement building practices within this research.
It encourages us to prioritize praxis---\ie, research activities and reflections directed at structural change---over proxy metrics of success---\eg, state-of-the-art performances and leaderboard rankings that do not translate to scientific or social progress).
And above all, it challenges the field to move beyond the restrictive view of IR research that emphasizes algorithmic advances as measurable on shared benchmarks and leaderboards to a more expansive view of IR research situated in a project of affecting social good and universal emancipation.

\paragraph{Emancipation research in other fields.}
To develop an emancipatory IR research agenda, we can take inspiration from other scholarly fields that have incorporated humanistic values and emancipatory aspirations.
\citet{young2021emancipation} surveyed the body of emancipation research in information science, and identified four components of emancipation prevalent in the literature: Agency (\ie, freedom to act), dialogue (\ie, freedom to express), inclusion (\ie, freedom to belong), and rationality (\ie, freedom to think).
\citet{wright2020envisioning} defines emancipatory social science as an intellectual enterprise seeking to ``generate scientific knowledge relevant to the collective project of challenging various forms of human oppression''.
Wright outlines a framework that enumerates three specific tasks for the field:
\begin{enumerate*}[label=(\roman*)]
    \item Systematic diagnosis and critique of the world as it exists,
    \item envisioning viable alternatives, and
    \item elaborating a theory of social transformation.
\end{enumerate*}
According to Wright, emancipatory social science is ``a theory of a journey from the present to a possible future: the diagnosis and critique of society tells us why we want to leave
the world in which we live; the theory of alternatives tells us where we want to go; and the theory of transformation tells us how to get from here to there – how to make viable alternatives,
achievable''.

In HCI, \citet{bardzell2016humanistic, bardzell2015humanistic} discuss emancipation under the heading of humanistic HCI, which they define as research and practices in the field that deploy humanistic epistemologies and methodologies.
They note that emancipatory HCI is ``a fundamental goal of virtually all humanistic HCI''.
Related, several strands of HCI research incorporate epistemologies, theories and practices that are anti-oppressive and emancipatory~\citep{smyth2014anti}, feminist~\citep{bardzell2010feminist, bardzell2011towards, bardzell2011feminism, bardzell2016feminist, bardzell2018utopias}, queer~\citep{light2011hci, klipphahn2024introduction}, postcolonial and decolonial~\citep{irani2010postcolonial, dourish2012ubicomp, sun2013critical, akama2016speculative, irani2016stories}, anti-racist~\citep{abebe2022anti}, anti-casteist~\citep{vaghela2022interrupting, vaghela2022caste}, anti-ableist~\citep{williams2021articulations, sum2024challenging}, post-capitalistic~\citep{feltwell2018grand, browne2022future}, and  anarchist~\citep{keyes2019human, linehan2014never}.

Similarly, in the fields of machine learning (ML) and artificial intelligence (AI) there are several works that call for perspectives and design interventions that are anti-oppressive and emancipatory~\citep{kane2021avoiding, saxena2023artificial}, anti-fascist~\citep{mcquillan2022anti}, decolonial~\citep{adams2021can, mohamed2020decolonial}, anti-casteist~\citep{kalyanakrishnan2018opportunities, sambasivan2021re}, and abolitionist~\citep{benjamin2019race, barabas2020beyond, earl2021towards, williams2023no}.
In data science, \citet{d2020data} and \citet{guyan2022queer} discuss how feminist and queer epistemologies can inform the field, and \citet{monroe2021emancipatory} adopted Wright's framework~\citep{wright2020envisioning} to propose an emancipatory agenda for data science to mitigate the harms to marginalized populations.

\paragraph{Towards emancipatory research in IR.}
\citet{belkin1976some} acknowledged information science's social responsibility nearly half a century ago.
But it was not until more recently that this perspective gained serious traction within the IR community, starting with the SWIRL~\citep{culpepper2018research} and the FACTS-IR~\citep{olteanu2021facts} workshops.
Subsequently, there has been a wide range of IR research on fairness~\citep{ekstrand2021fairness}, explainability~\citep{anand2022explainable}, and addressing misinformation~\citep{zhou2020survey} among other societally-motivated topics.
However, the field has been largely reluctant to acknowledge the saliently political nature of this work leaving the underlying colonial, cisheteropatriarchal, and capitalist values that has (and continue to) critically shape the field of IR unchallenged.
Even IR conference tracks dedicated to research that tries to affect social good have often side-stepped the deeply sociopolitical question of defining what constitutes social good~\citep{mitra2025ir}.

This has led to several recent calls to explicate and critically examine the norms and values~\citep{vrijenhoek2023report, trippas2025report} shaping the field as well as the sociotechnical futures that the IR community wants to realize through their research~\citep{mitra2025search, azzopardi2024report}; while also reasserting the interdisciplinary roots of IR~\citep{zangerle2025beyond}.
This year, the ECIR IR-for-Good track's call for papers\footnote{\url{https://ecir2026.eu/calls/call-for-ir-for-good-papers}} explicitly define IR-for-Good as ``\emph{IR research and practices that contribute towards realizing more equitable, emancipatory, and sustainable futures}''~\citep{mitra2025ir}, and updated the list topics relevant to the track to shift the emphasis from desired attributes and interventions in IR systems (\eg, fairness and accessibility)---that were the norm in previous years\footnote{\url{https://ecir2025.eu/call-for-ir-for-good-papers/}}---to desired justice-oriented real-world outcomes (\eg, racial, disability, and sexuality justice) that IR research should try to realize.
They also require papers to include explicit discussion on their theory of change.

Several of these developments---\ie,~\citet{trippas2025report, azzopardi2024report, mitra2025ir}---were prompted by conversations based on an earlier preprint of this current manuscript.
In this work, we argue that moving forward the IR community should adopt emancipation as an explicit objective for the field. 
\section{Practices, projects, and provocations}
\label{sec:how}

\begin{figure*}
    \centering
    \includegraphics[width=1.2\linewidth]{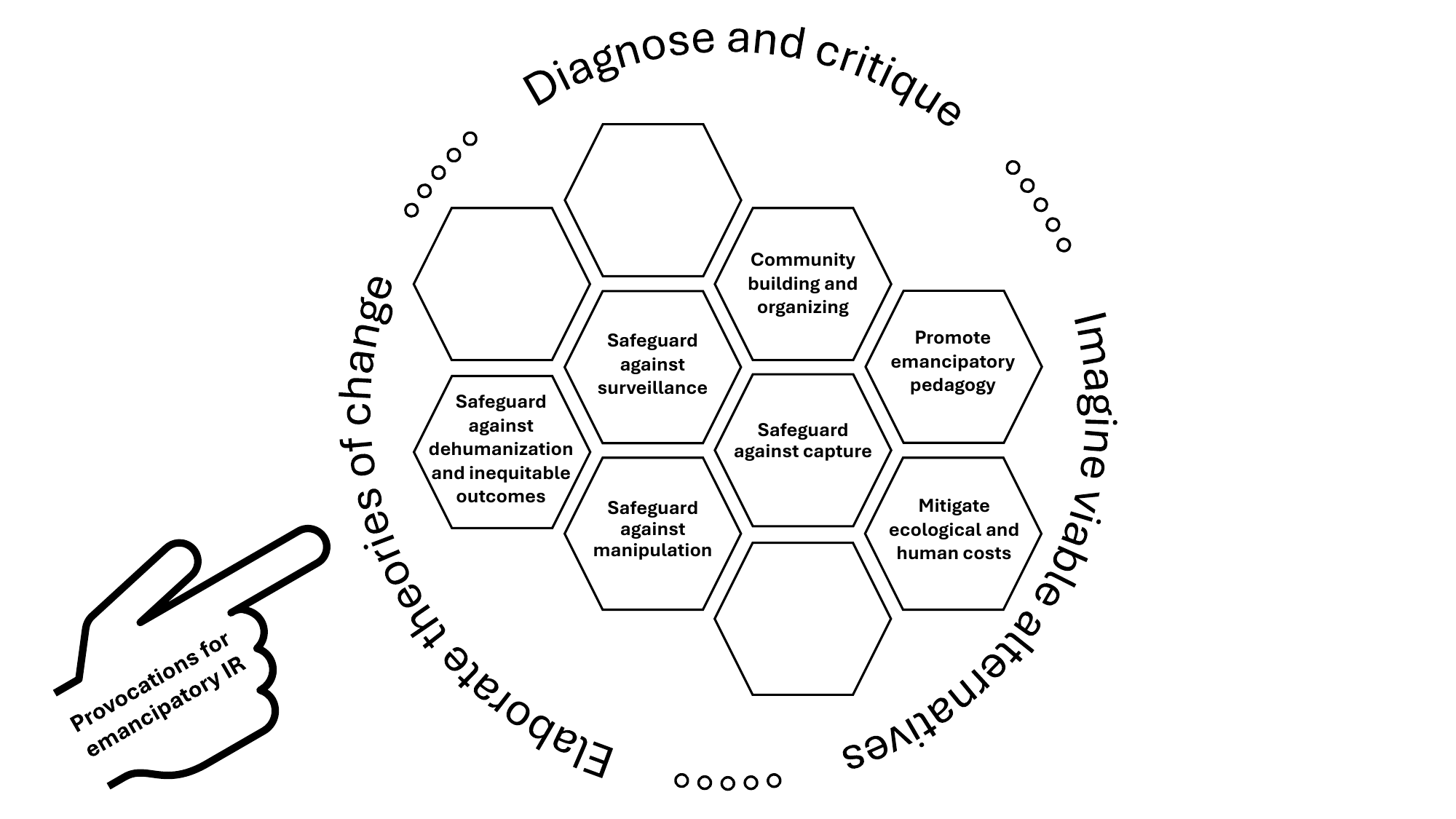}
    \caption{Illustrating the practices, projects, and provocations of emancipatory IR.
    The hexagons at the center represent several potential projects.
    The empty hexagons indicate that the set of mentioned projects are not complete, and the community should over time identify new ones and shape existing ones, as appropriate.
    Surrounding the projects is the framework of practices of emancipatory IR that are relevant to all projects in this area, consisting of (i) diagnose and critique, (ii) imagine viable alternative futures, and (iii) elaborate theories of change.
    Provocations to spark new ideas and research directions are depicted as a nudging hand.}
    \label{fig:practices-projects-provocations}
\end{figure*}

To develop an emancipatory research agenda for IR, we speculate about the practices that the community can adopt, enumerate some of the projects that the field should undertake, and discuss provocations to spark new ideas and directions for research.
However, the list of practices, projects, and provocations discussed here are neither immutable nor complete.
It is ultimately the responsibility of the community engaged in emancipatory research to regularly reflect on, produce evidence-based critique, challenge, and reshape its agenda to affect tangible progress in our collective struggles against dehumanization, oppression, and marginalization.
It may be worth reemphasizing at this stage that due to the saliently sociotechnical nature of the challenges in this area, the practices, projects, and provocations too will rarely be purely technological or reside strictly in the domain of computing, and must be informed by perspectives from outside of the field of computer science and the lived experiences of marginalized peoples.
We illustrate the practices, projects, and provocations described in this section in Figure~\ref{fig:practices-projects-provocations}.

\subsection{Practices of emancipatory IR}
\label{sec:how-practices}
We adapt Wright's emancipatory social science framework~\citep{wright2020envisioning} to develop a framework for the practices of emancipatory IR.
Like Wright's, our framework consists of diagnosing and critiquing, imagining viable alternative futures, and elaborating our theories of change.

\paragraph{Diagnose and critique.}
The starting point for emancipatory IR research is identifying the ways in which existing IR methods and systems, the cost of IR research and development, and the arrangements within the IR community may contribute systemic harms to peoples, act contrary to aspirations of social and political justice, and impede emancipatory struggles of the peoples.
These harms may include:
\begin{itemize}
    \item Suppressing voices of the oppressed and marginalized
    \item Enabling surveillance and public opinion manipulation
    \item Imposing representational harms through derogatory portrayal, stereotyping, or erasure in presentation of retrieved results
    \item Imposing allocative harms to groups through under-exposure of their content in retrieved results or under-exposing groups to socioeconomic opportunities
    \item Exploitation of ecological resources and labor for IR research and development that mirror racial capitalism and coloniality
    \item Capture of IR research by the military–industrial complex that pushes the field to disproportionately prioritize economic and military interests over concerns of knowledge production, public health education, and information literacy, and produce technologies that serve as tools of oppression (\eg, for surveillance)
    \item Concentration of extreme power and wealth among few individuals and institutions in ownership of popular information access platforms
    \item Lack of representation and diversity within the IR community leading to propping up colonial, cisheteropatriarchal, and capitalist values and erasure of feminist, queer, decolonial, anti-racist, anti-casteist, anti-ableist, and abolitionist perspectives
\end{itemize}

\noindent
Diagnose and critique must extend to the policy and regulatory landscape, such as analyzing the European Union’s General
Data Protection Regulation (GDPR) ~\citep{regulation2016regulation} and Digital Services Act~\citep{european2022regulation} in the context of our emancipatory goals which we identify as critical future work.

\paragraph{Imagine viable alternative futures.}
The second practice of emancipatory IR is to imagine and develop desirable, viable, and achievable alternatives to our IR technologies and ways of organizing our communities with the aim of addressing the harms and injustices identified in the diagnosis and critique stage.
Previous work~\citep{mitra2025search} have argued that the IR community should explicitly articulate the sociotechnical imaginaries~\citep{jasanoff2009containing, jasanoff2015dreamscapes} that influence the design and development of IR technologies and platforms; and enquires ``what would IR systems look like if designed for futures informed by feminist, queer, decolonial, anti-racist, anti-casteist, anti-ableist, and abolitionist thoughts, and if the focus of IR research was not to prop up colonial cisheteropatriarchal capitalist structures but to dismantle them?''
This is a call for us to imagine post-oppressive worlds, and the role of IR both in that world and in the journey that leads us there.

We intentionally do not speculate about any specific futures in this paper.
To satisfy that ask, IR needs spaces and processes for design futuring~\citep{fry2009design}.
The goal of futuring here is not just to envision a plurality of alternative utopian and protopian futures, but also to build shared understanding, aspirations, and commitment towards emancipatory outcomes through that participatory process.
After all, the act of exercising one's imagination itself can be liberating, or as \citet{benjamin2024imagination} calls it ``an invitation to rid our mental and social structures from the tyranny of dominant imaginaries''.

\paragraph{Elaborate theories of change.}
Our final practice concerns with the development of theories of change~\citep{weiss1995nothing, brest2010power, taplin2012theory, wiki:TheoryOfChange} that makes explicit how we aim to realize the alternative futures in the face of current realities and how our own work contributes towards those goals.
This should be grounded in theories of power and co-developed with cross-disciplinary scholars, legal and policy experts, activists, artists, and others to realize real structural changes.
This involves both developing new research agendas as well as experimenting with and realizing new arrangements within the IR community and critically reflecting on our relationships with other institutions (\eg, industry and government).
To provide concrete examples, we next discuss potential projects that we believe emancipatory IR should invest in.

\subsection{Emancipatory projects in IR}
\label{sec:how-projects}
Next, we enumerate some of the potential projects relevant to emancipatory IR.
Such examples are useful for readers to get a more concrete sense of the aspirations and type of work relevant to this field, and can also serve as a starting point for developing a more thorough research agenda.
Ultimately, the direction and what specific project we invest in must be decided and iterated on over time by those who see themselves as part of this community, which we must ensure includes more than just IR researchers and technologists as we emphasized previously.

\paragraph{Safeguard against capture.}
A central concern of emancipatory IR should be to ensure that our information access platforms cannot be easily captured by few privileged individuals and institutions or authoritarian regimes, and also to ensure that access to information services cannot easily be blocked by authoritarian regimes in times of protests to suppress dissent.
These concerns and initiatives are timely in the face of the 2024 edition of the World Economic Forum’s Global Risks Report~\citep{wef2024world} that calls out ``technological power concentration'' as one of the top global risks for the upcoming decade, and broader growing concerns about the alignment of tech industry with authoritarianism and fascism~\citep{kang2025big, gebru2023eugenics, varoufakis2024technofeudalism, lafrance2024rise, duran2024, akbari2025big}.
In the social media landscape, there have been recent announcements~\citep{mastodon2025people, bsky2025free} of formation of new non-profit foundations for managing key social media ecosystems and platforms, specifically with the goal to safeguard against billionaires' control over our digital public squares.
In IR, there has also been recent investments to create open infrastructures for internet search, such as the Open Web Search project.\footnote{\url{https://openwebsearch.eu/}}

While decentralization of IR infrastructure and federation may be critical for protecting against capture, it is also important to acknowledge that in a world shaped by white supremacy, colonialism, patriarchy, cisheteronormativity, neuronormativity, ableism, and casteism as organizing sociopolitical forces, the decentralization of the technological infrastructures in themselves do not constitute sufficient conditions for creating safe and welcoming online spaces for historically marginalized peoples nor to ensure these platforms align with our emancipatory aspirations~\citep{hendrix2022whiteness}.
Instead, the challenges that we face require both decentralized technology infrastructure, as well as social innovations in the formation of new global governance structures for the ownership, management, and moderation of these platforms to ensure that the voices of the oppressed and marginalized are uplifted, not suppressed.

\paragraph{Safeguard against surveillance.}
IR applications---such as web search, recommender systems, and targeted digital advertising---have been major drivers of surveillance capitalism~\citep{zuboff2019age}.
An important commitment IR can make towards anti-oppression is to ensure that the field is neither creating the technologies and infrastructures for surveillance, nor creating additional demands for such infrastructures out of commercial interests.
User behavior data have been the ``secret'' sauce for many successful commercial IR platforms.
For example, in search, user interactions with search result pages are continuously logged and modeled at incredible scale to provide better search relevance to customers.
It is therefore not surprising that modeling of user clicks~\citep{chuklin2015click} and even cursor movements~\citep{diaz2013robust} have historically garnered substantial interest in IR research.
However, this creates demands for unscrupulous surveillance of users on these platforms.
In response, different regulatory safeguards have been proposed, including notably the principles of data minimization defined in Article 5 of the GDPR~\citep{regulation2016regulation}.

To safeguard against surveillance, the IR community must imagine alternative ways of building its systems that does not necessitate such large-scale data collection to simply make these platforms usable.
We must both push the boundaries of effective retrieval methods in the absence of user behavior data \emph{and} develop new protocols for collection, storage, and application of user data that implements stricter safeguards for user privacy and require stricter and more meaningful user consent for access by the platforms.
We need to explore both technological innovations (such as, homomorphic encryption) as well policy safeguards and changing social norms around data stewardship and transparency of how institutions collect, store, and use data, and implement concrete safeguards to ensure that authoritarian governments and their institutions cannot get access to this data which may enable harassment and intimidation of activists, journalists, and marginalized populations.
Lastly, we should be wary of potential dual-use of IR technologies---such as the impact of improved cross-lingual search on surveillance of foreign individuals, institutions, and governments---and be critical of and refuse funding for such research that comes from governments and security agencies.

\paragraph{Safeguard against manipulation.}
Misinformation and disinformation pose serious threats to democracies worldwide~\citep{lewandowsky2023misinformation, ecker2024misinformation, lewandowsky2024truth}.
The 2024 edition of the World Economic Forum’s Global Risks Report~\citep{wef2024world} ranks misinformation and disinformation as the top global risk for the subsequent two years.
A study spanning six years, 26 countries, and several election periods find that political misinformation is particularly salient in current wave of radical right populism and its opposition to liberal democratic values and institutions~\citep{tornberg2025parties}.
Curtailing spread of online misinformation and disinformation that aims to dehumanize immigrants, the colonized, the poor, trans and nonbinary, and racially marginalized peoples, is critical to our social justice efforts and our effort to build solidarity with each other.

Beyond misinformation, we must also pay close attention to new vectors of public manipulation that may become plausible in near future based on our current technological trends.
A particularly worrying development in this area is the recent emergence of ``persuasive AI''~\citep{burtell2023artificial, carroll2023characterizing, park2023ai, el2024mechanism}.
The massive trove of detailed data on user behavior and preferences combined with the capabilities of generative AI to produce persuasive language and visualizations could create tools of mass manipulation and pose serious risks to functioning of global democracies~\citep{mitra2024sociotechnical}.
Imagine every time you searched online or accessed information via your digital assistant, the information was presented to you exactly in the form most likely to alter your consumer preferences or political opinions.
Or, consider AI-generated digital characters in ads and videos that appropriate marginalized identities to say or act in ways that real members of that community may be strongly opposed to---a new form of ``Digital Blackface''~\citep{johnson2023generative}.
Emancipatory IR researchers must engage in evidence-based critique of such technology development and push for policies and community norms that push back against research in support of such applications within the field before more such applications of AI materialize and are normalized.

\paragraph{Safeguard against dehumanization and inequitable outcomes.}
Emancipatory IR should concern itself with ensuring that IR systems do not lead to representational harms (\eg, denigration, stereotyping, and erasure) of historically marginalized peoples and inequitable outcomes which may further reinscribe historical sociopolitical and economic oppression and exploitation.
Over the last several years, there has been several works on fairness in IR.
These have typically focused on fairness of quality-of-service---\eg,~\citep{mehrotra2017auditing, mehrotra2018towards, neophytou2022revisiting, wu2024towards}---and fairness of exposure---\eg,~\citep{asia:equity-of-attention, singh2018fairness, singh:fair-pg-rank, diaz2020evaluating, zehlike2020reducing, PatroBGGC20FairRec, wu2022joint}.
However, there is less clarity on how much this body of fairness research has practically influenced material changes in popular IR platforms~\citep{mitra2025search}.
Fairness research in the emancipatory context must move beyond just formalization and theorization of fairness; it must recognize and be grounded in the understanding of historical injustices and structural violence, and must ensure that this research opposes dehumanization and material inequities in the real world.
This includes also contending with the politics of classification~\citep{crawford2021atlas} and critiquing practices in fairness research, such as in context of gender fairness~\citep{pinney2023much}, that may themselves perpetuate further harm.

We must also hold institutions that own and operate IR platforms accountable to measurable equitable and humanistic outcomes.
If an institution invests in fairness research but does not operationalize them in their products, then that research only serves to launder institutional reputation and is counter to our real emancipatory goals.
To hold institutions accountable, our research agenda should also include challenges of auditing IR platforms and also explore other regulatory, sociopolitical, and technological mechanisms of accountability that are grounded in the recognition of the structural power inequities that make it challenging to do so.

\paragraph{Promote emancipatory pedagogy.}
Emancipatory IR must abandon any false notions of neutrality between the oppressor and the oppressed, and develop frameworks situated in theories of power to uplift marginalized voices and make spaces for emancipatory pedagogy.
It must expressly refuse to serve as a soapbox for the powerful and wealthy.
To make this further challenging, we must do so without concentrating the power to define oppressor-oppressed relations in the hands of platform owners.
Can we, for example, learn from Freire's~\citep{freire2020pedagogy} anti-oppressive pedagogy to imagine IR systems that do not merely retrieve but provide spaces for dialogical interactions between information seekers?
Can we build communities of experts who can provide context from critical scholarship to inform search results and aid in our collective sense-making and solidarity-building on topics relevant to our emancipatory struggles?
What if we go beyond fixing under-representation of marginalized groups in, say, image search results for occupational roles and reclaim those digital spaces as sites of resistance and emancipatory pedagogy~\citep{mitra2025search}? 

Emancipatory IR must also critically interrogate the relationship between information access platforms and the publishers and news media.
\citet{kim2024strategies} find media indoctrination as one of the strongest predictors of autocratic survival.
In light of these dynamics, we must reexamine how platforms can support local newsrooms that often do critical work in reporting on issues relevant to marginalized populations.
In this context, support must include both ensuring these publishers get exposure in search results as well as innovations in platform business models that can support local newsrooms financially, such as through revenue sharing.

\paragraph{Visibilize and mitigate ecological and human costs of IR research and development.}
We must also be concerned of the cost of doing IR research and operating IR platforms.
This includes ecological costs~\citep{scells2022reduce, zuccon2023beyond}, such as energy and water consumption and harmful emissions, as well as the appropriation of data labor and impact on the wellbeing of workers.
\citet{belkhir2018assessing} estimate that Information and Communications Technology industry on the
whole will account for 14\% of global emissions by 2040.
In the US, data centers
are projected to consume around 6\% of the total national electricity by
2026~\citep{halper2024amid}.
Appropriation of data labor includes both uncompensated appropriation of works by writer, artists, and programmers~\citep{cohan2023ai, coldewey2023thousands, vincent2023ai, vincent2022lawsuit, shrivastava2023openai, burke2023biggest, burke2024generative, vincent2021github, vincent2020dont, appel2023generative, marr2023generative, chesterman2024good, chayka2023ai, gertner2023wikipedia, vincent2023chatgpt} and under-paid crowdwork for data labeling~\citep{perrigo2023exclusive, williams2022exploited, tan2023behind, altenried2020platform, hao2022ai, xiang2023openai, hao2023cleaning}.
The latter mirrors historical patterns of global labor exploitation and racial colonial capitalism~\citep{hao2022artificial, birhane2020algorithmic, o2023heart, klein2023ai, couldry2019data, muldoon2023artificial, tacheva2023ai}, and have been associated with severe mental health cases among crowdworkers~\citep{booth2024more, hao2023cleaning}.
Other concerns include land use for data centers and how that affects communities living on those lands and indigenous sovereignty~\citep{rubayita2025alberta, verma2024shadows}.

Emancipatory IR must invest in accurate accounting, reporting, and reduction of the ecological and human cost of our research.
This includes the cost of training models and experimentation, the cost of human annotations, and the cost of air travel for conference attendance, among others.
It includes a commitment towards not pursuing research directions whose cost include accelerating climate change and broader adoption of virtual conferencing to minimize air travel.
It also includes the recognition that these costs cannot just be minimized by building more efficient models---\eg, because of Jevons paradox\footnote{\url{https://en.wikipedia.org/wiki/Jevons_paradox}}---or automated labeling methods, and that the solution as always is at the intersection of the social, the political, the economic, and the technological.

\paragraph{Community building and organizing.}
The work of building an emancipatory research agenda must start with building an emancipatory research community.
We need to create safe spaces where researchers from the IR community can engage and meaningfully interact with a diverse set of scholars, designers, legal and policy experts, activists, and artists to co-develop an agenda for this field.
We must reflect on the current social and organizational arrangements within the field and our relationships with other institutions, such as industry and government.
We must take on the difficult challenge of critiquing the funding mechanisms that enable IR research and its role in privileging certain areas of research explorations and technology development that are counter to our emancipatory aspirations and social justice.
We must reimagine these arrangements, build counter-structures, and establish new means of funding emancipatory IR research that can sustain itself in the face of attack from the privileged, powerful, and wealthy.
We must nurture a culture in our community that recognizes and encourages direct actions to challenge power, such as demonstrating and whistleblowing~\citep{hicks2025history}, and create support mechanisms to protect individuals from reprisal from those whose power and visions we want to challenge.
We must ensure that our community is inclusive and representative of marginalized peoples and that we view their lived experiences with structural oppression as critical to informing research in our field.
And we should view emancipatory IR as situated in broader movement building against marginalization, dehumanization, and oppression.

\subsection{Provocations}
\label{sec:how-provocations}
Emancipatory IR must leverage instruments of provocation to challenge our thinking and help us reimagine radically alternative futures.
Expanding on previous call~\citep{mitra2025search} to safeguard the IR community from falling victim to crises of imagination~\citep{haiven2014crises}, we posit that the inception of emancipatory IR must not be in a vacuum, but must be informed by emancipatory ideologies borne of our historical and ongoing struggles, including feminist, queer, decolonial, anti-racist, anti-casteist, anti-ableist, and abolitionist thoughts.
Our challenge is not only to imagine radically new emancipatory futures, but also to put into practice those desired values and build prefigurative counter-structures both in the form of new technological designs and sociopolitical rearrangements within our communities.
The task to change technology is inextricably linked to the task of changing ourselves and our own social arrangements.
The practice of reflexivity becomes important in this context to articulate (and critique if necessary) our own values, positionality, and incentives, as well as our relationships with institutions and social structures that we in turn intend to change.
It is in this process that we desperately need to engage scholars outside of computing, learn from generations of activists and movement organizers who have been engaged in emancipatory struggles, and let ourselves be influenced by the artists and writers that dare to imagine radically free futures and humanistic social structures.

\section{Discussion}
\label{sec:discussion}

\subsection{The \emph{emancipatory} in emancipatory IR}
\label{sec:discussion-emancipatory}
For emancipatory IR to be authentically emancipatory, this work must not be simply limited to an intellectual exercise.
Our scholarship must be a piece of a broader movement building praxis grounded in ongoing struggles in our world against marginalization, dehumanization, and oppression.
Our critiques must be followed by a tangible push for dismantling oppressive structures and building counter-structures in its place.
Our theories must be put into practice and tested in the real world, and our success measured in concrete emancipatory outcomes.
Access to information is critical to our collective sense-making of our place in this world and our relations with it, and it is exactly why throughout history, from print media to present-day social networks, authoritarian forces have tried to control what and how information is disseminated.
Emancipatory IR stands in solidarity with the oppressed and their struggles, and accepts the challenge of building the information infrastructure for resistance.

\subsection{The \emph{IR} in emancipatory IR}
\label{sec:discussion-IR}
But what makes emancipatory IR truly \emph{IR}?
How are questions of emancipation relevant to this field of computing whose research is typically concerned with questions of ranking, evaluation, efficient system design, language modeling, human information interaction, and other similar topics?
To this we respond by challenging the reader to reflect on how each of these important concerns in IR intersect with the exemplar projects we discussed in Section~\ref{sec:how-projects}.
What new research questions emerge in context of indexing and ranking when our IR platforms are federated and decentralized?
What new research challenges emerge in human information interaction if our goal is to promote emancipatory pedagogy?
How do we mitigate the harms from appropriation of data labor that is so very prevalent in our modern-day language modeling approaches?
How do we reimagine efficiency and evaluation when our aspirations go beyond fetching information artifacts at lightning speed to encouraging deeper collective understanding and new knowledge production that are key to our social progress?
With these questions our aim is to challenge the field to abandon a more restrictive view of what IR research is which may limit itself to algorithmic advances as measurable on shared benchmarks, and adopt a more expansive view of IR that aims to situate information access as a vehicle for social good.
These aspirations are also reflected in the recent SWIRL 2025 report~\citep{trippas2025report} that calls for centering emancipatory values in IR research.

\subsection{Where do we begin?}
\label{sec:discussion-getstarted}
Bootstrapping an emancipatory IR research agenda will be challenging.
The process requires existing IR researchers to expand their horizons of concerns and expertise, bring in others from outside of the field experienced in emancipatory struggles, and consider new assemblages of both its technologies and sub-communities.
While there may be several potential starting points for this process, we recommend that the work should begin by pulling together a seed community interested in this research direction.
Towards that goal, one proposal would be to organize community-building workshops whose structure is informed by the practices, projects, and provocations discussed in Section~\ref{sec:how}.
A sketch of such a workshop may look as follows:
\begin{enumerate}
    \item A series of talks covering critiques and provocations relevant to emancipatory IR.
    \item A design futuring session where participants are encouraged to radically reimagine IR systems and how they should be embedded in our society and our daily lives.
    \item Following the futuring exercise, participants sketch out paths to our desired futures and identify concrete challenges and initiatives that the emancipatory IR community should take on.
    Some of these may align with the projects enumerated in Section~\ref{sec:how-projects} or may identify the need for additional projects that we have not considered here.
    \item Finally, the participants must discuss funding considerations to sustain our research agenda, as well as steps that we can take to protect our community members from reprisal from whose power our research aims to challenge.
\end{enumerate}

The goal of such workshops would be to build trust, solidarity, and a shared understanding of our aspirations and plans within the seed community.
It is also our aim with these workshops to challenge the seemingly unassailable wall of colonial capitalist values that have arrested our present visions of society and technologies.
We hope such a workshop can be successful in finding the cracks in the wall through which we can collectively peep out and free ourselves to imagine a plurality of social arrangements that are possible in which we can all flourish free from oppression.
\section{Conclusion}
\label{sec:conclusion}
The field of IR has undergone seismic shifts over the last decade largely instigated by rapid developments in deep learning~\citep{mitra2018introduction} and more recently generative AI~\citep{white2025information}.
Between trying to stay updated in the face of daily firehoses of new AI publications and chasing the coveted badges of ``state-of-the-art'' it can be easy as an IR researcher to overlook the pressing needs of society and envision only the possibilities as conjured by the latest AI news cycle.
If we are to address the multitude of crises facing our world today---spanning the ecological, the social, the political, and the economic spectrums---IR must undergo another seismic shift over the next decade.
This piece is a provocation for the IR community to situate itself in the generational emancipatory struggles and stand in resistance to the growing epidemic of authoritarianism and fascism through the practices of making and unmaking of information access technologies.
It is also an invitation to scholars from other fields as well as to social movement organizers and artists to join us and be part of this shared struggle.
And as IR researchers, we must remember that ``it is vital that we approach these spaces with curiosity and humility; in recognition of our own incomplete understanding of the world; open to change and be changed by these encounters''~\citep{mitra2025search}.

A central thesis underlying this work is that the emancipatory struggles in the real world and one that we are encouraging the IR community to acknowledge and center within our research are one and the same.
So, the work of emancipatory IR and the work of movement building for social justice and emancipation outside of IR must co-construct the conditions for dismantling oppressive regimes, such as those of colonialism, cisheteropatriarchy, and capitalism.
While focused on IR, it is also a challenge to the broader computing and technology community to reimagine our work to not serve the privileged few by further concentrating immense power and wealth, but as a liberatory force that connects us and humanizes us universally.
Technology after all \emph{is} a project of world-making.
So, what kind of a world are \emph{we} collectively going to build together?

\section*{Positionality statement}
I, the author of this paper, spent most of my career at a large technology corporation in the global north.
However, the perspectives presented in this work is intended to challenge Big Tech and global north's view of technology and our collective futures.
Outside of information retrieval research, I participate in social movement spaces and Southasian anti-fascist organizing.
I stand in solidarity with the peoples of Palestine, Sudan, Congo, Rohingya, and other groups who are being subjected to unimaginable oppression and ongoing genocide.
I also stand in solidarity with the people of Kashmir in their emancipatory struggles against state violence.
These ongoing atrocities and cases of systemic oppression significantly informed and influenced my work.

\acks{
The author gratefully acknowledges feedback from Ida Larsen-Ledet on the various drafts of this paper.
No external funding was received in support of this work.
}

\vskip 0.2in
\bibliography{references}
\end{document}